\documentstyle[psfig]{caosp}

\def\Fe2{\hbox{{\rm Fe}~{$\scriptstyle {\rm II}$}}}
\def\Cr2{\hbox{{\rm Cr}~{$\scriptstyle {\rm II}$}}}
\def\kms{\,km\,s$^{-1}$}
\def\degr{\deg}
\begin{document}


\title{Doppler Imaging of Ap Stars}

\author { R. Kuschnig}
\institute{Institut f\"ur Astronomie, 
Universit\"at Wien, T\"urkenschanzstr. 17, A-1180 Vienna, Austria}

\maketitle

\begin{abstract}
Doppler imaging, a technique which inverts spectral line profile variations
of an Ap star into a two-dimensional abundance maps, provides new observational
constraints on diffusion mechanism in the presence of a global magnetic
field. A programme is presented here with the aim to obtain abundance 
distributions of at least five elements on each star, in order to
study how different diffusion processes act under influence of a stellar
magnetic field.
The importance of this multi-element approach is demonstrated, by presenting
the abundance maps of helium, magnesium, silicon, chromium and iron
for the magnetic B9pSi star CU Vriginis. 

\end{abstract}


\section{Motivation}

The scientific goal of applying the Doppler imaging technique to Ap stars, is to
derive observational constraints on the diffusion mechanism in the presence of a
stellar magnetic field. Theoretical studies (Michaud et al, 1981; M\'egessier,
1984) which predict that certain elements raise, sink or move horizontally in
magnetic stellar atmospheres need to be verified. This can only be achieved if an
unambiguous correlation between the elemental abundance structures and the magnetic 
field geometry is obtained. However, two major problems are responsible for
difficulties in reaching that goal. 

Firstly, Doppler imaging can (presently) only be applied
to stars which have a fairly weak magnetic field and hence only the 
variation of the integrated longitudinal field strength (effective field,
$B_{\rm eff}$) is 
usually determined. This sets only weak limits to the `real' magnetic geometry.
For objects with stronger fields the spectral line profiles are severely influenced
by Zeeman broadening and are not mainly caused by abundance variations.
This limitation can be avoided by obtaining all four Stokes parameters as a
function of the stellar rotational phase. Then Zeeman Doppler imaging codes
(Donati, 1995) can be applied to calculate the magnetic field strength and surface
orientation in combination with the abundance distribution. But these
measurements of the Stokes parameters are not yet available, though strong
efforts are made to get them from polarimetry. 

Secondly, in the past, only a few elements were mapped, mostly 
silicon, rarely iron and chromium, not even all of them for each object. 
Furthermore, these maps were based on inversions from single `unblended' 
spectral lines, which are difficult to find, especially in cooler Ap stars. 
This introduced uncertainties leading to a questionable reliability of the 
derived maps and weakening their relevance to theoretical modelling.

Considering this situation, we focused our programme on the production of 
abundance maps of up to five and more elements on the surface of
each of our target stars. A brief report on the stars and the progress of our 
project is given hereafter, followed by the presentation of results obtained for the
magnetic Ap star CU Virginis.

\section{Multi-element Doppler imaging programme}
As pointed out before, one crucial limitation of Doppler imaging in the past  
was that only a few elements were mapped, mostly silicon and
occasionally iron and chromium.  Furthermore, the number of stars studied
was not sufficient to draw general conclusions by relating the abundance maps
to the approximate magnetic field structure, thereby providing new constraints
on theory (Hatzes 1995). In particular, the abundance maps of elements
which are pushed upwards by radiatively driven diffusion should look severely 
different from those of elements that have a tendency to sink. 

\begin{table}
\begin{center}
\begin{tabular}{||l|l|l|l||}
\hline\hline
Star & HD & Type & Elements \\
\hline\hline
$\iota$ Cas & 15089 & A5pSr & Mg,Ti,Cr,Fe \\
$\theta$ Aur & 40312A & A0pSi & Mg,Si,Ti,Cr,Mn,Fe \\
$\epsilon$ UMa & 112185 & A1pCrEuMn & O,Mg,Si,Ti,Cr,Mn,Fe \\
CU Vir & 124224 & B9pSi & He,Mg,Si,Cr,Fe \\
BP Boo & 140728 & A0pSiCr & Mg,Si,Ti,Cr,Fe \\
 &	153882 & A1pCrFe & Cr,Fe \\
$\phi$ Dra & 170000 & A0pSi & O,Mg,Si,Ti,Cr,Fe \\
ET And & 219749 & B9pSi & He,Mg,Si,Ti,Cr,Fe \\
\hline\hline
\end{tabular}
\end{center}
\caption{Ap stars and the elements which were mapped}
\end{table}

Table 1 comprises the present results of our programme.
For five more stars, observations were made, which still need to be
analyzed or supplemented for missing phases. Presentations of
surface images for $\iota$ Cas (Kuschnig at al.) and $\epsilon$ UMa 
(L\"uftinger et al.) are part of these procceedings. Here,
the results for the B9 silicon star CU Vir are presented and 
discussed. 

\section{Five elements on the surface of CU Virginis}
The spectra of CU Virginis were obtained at
Observatoire de Haute-Provence using the AUR\'ELIE spectrograph, in
1994 and 1995.  The spectral resolution of these data is about 
$2\times 10^4$, and the signal-to-noise ratio is typically 200:1.
The abundance maps were calculated using the Doppler imaging 
technique described by Piskunov \& Rice (1993). 
The input data for mapping are given in Table 2.

\begin{table}
\begin{center}
\begin{tabular}{||l|c||}
\hline\hline
 & CU Vir\\
\hline
Ephemeris & $243178.9025$ \\
	 &  $+ 0\fd52070308 \cdot {\rm E}$ \\
$v\sin i$ & 160~\kms  \\
Inclination &  $30\degr$  \\
$T_{eff}$ & 13000~K \\
$\log g$  & 4.0 \\
\hline\hline
\end{tabular}
\end{center}
\caption{Input data for the abundance Doppler imaging of CU Vir}
\end{table}

The effective field variation of CU Vir was measured by Borra \& Landstreet (1980)
and can be modelled to first order by a decentered dipole geometry (Hatzes, 1995).

The helium map (Fig. 1) is characterized by a dominant spot which appears to
be at the approximate position of the positive magnetic pole (phase 0.5).
Only in the central part of this spot does helium reach the
solar abundance; on all other parts of the stellar surface, it is depleted by 
about 1.5 dex. This result confirms theoretical predictions by Vauclair et al. (1991)
which have been obtained by modelling helium abundance in main sequence 
magnetic stars by introducing a weak wind of ionized metals at
the magnetic poles, where the field lines are vertical.

In total contrast to that, silicon (Fig. 2) is strongly depleted in the helium 
spot but overabundant on the remaining visible surface. In these regions the
magnetic field lines are mainly horizontal and as predicted by e.g.
M\'egessier (1994), silicon accumulates at the magnetic equator band.
Chromium and iron have surface structures very similar to that of 
silicon, but both elements are less enhanced compared to their solar abundances.

However, the magnesium (Fig. 3) distribution differs much from
that of the other metals. The main feature is a ring centred at the
helium spot (magnetic pole), but with an the overall abundance deficiency
of about 1 dex and much less contrast than all other elements.

\begin{figure}
\centerline{\psfig{figure=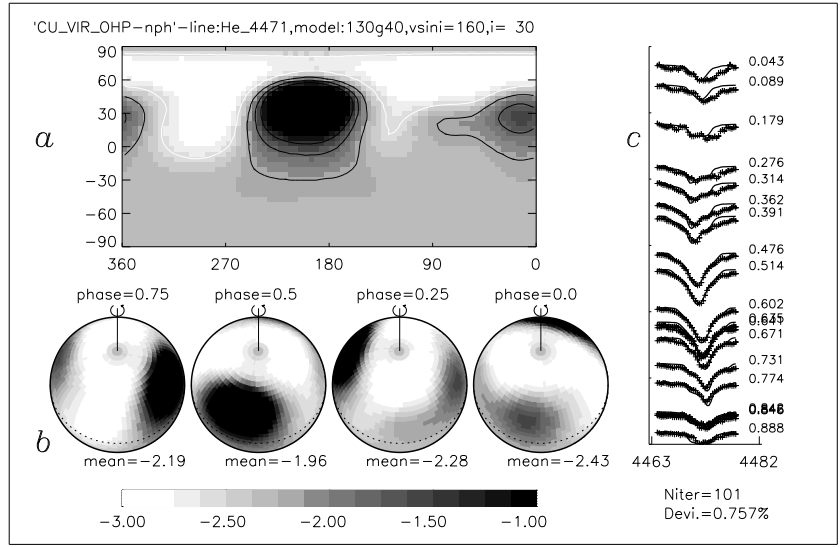,height=7.2cm}}
\caption{Helium abundance distribution of CU Vir obtained 
from the He\,{\sc i} 4471\AA\ blend.}
\end{figure}

\begin{figure}
\centerline{\psfig{figure=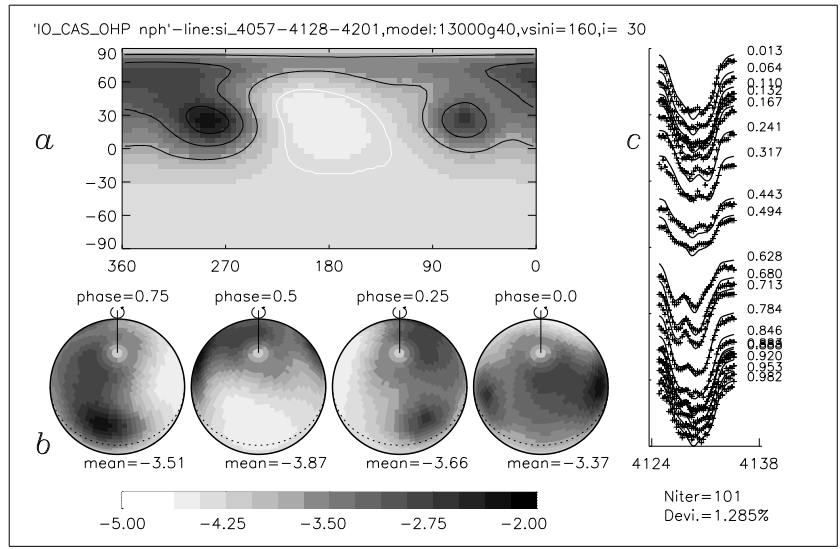,height=7.2cm}}
\caption{Silicon abundance distribution of CU Vir. More than
20 Si\,{\sc ii} lines in three separate wavelength regions (4057\AA ,
4128\AA , 4201\AA ) have been used to calculate this map.}
\end{figure}

\begin{figure}
\centerline{\psfig{figure=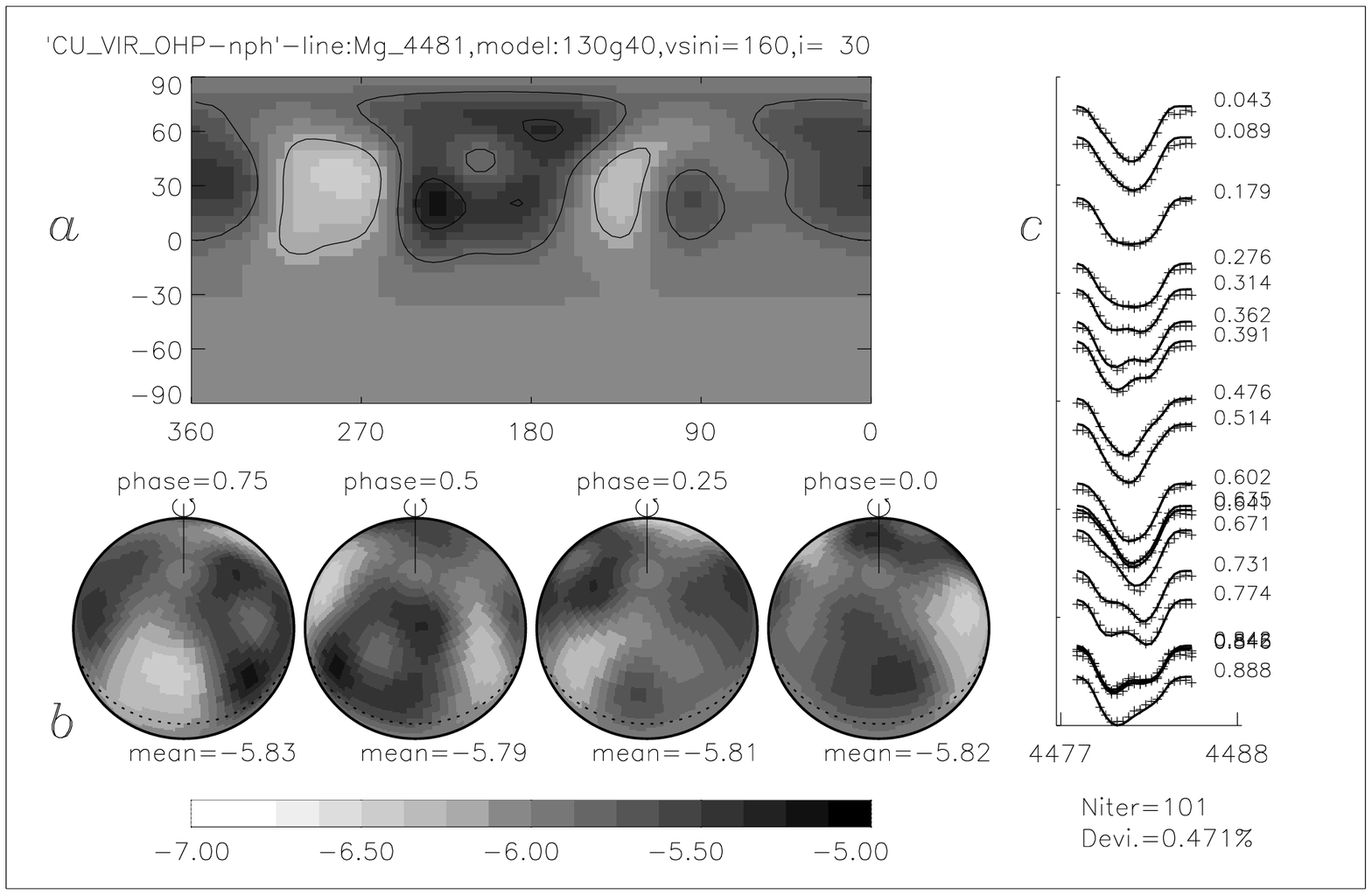,height=7.2cm}}
\caption{Magnesium abundance distribution of CU Vir obtained
for the Mg\,{\sc ii} 4481\AA\ lines.}
\end{figure}

The helium spot distribution of CU Vir was already found by 
Hiesberger et al. (1995), on the basis of data obtained in 1980, which
is an evidence for the high reliability of this result. 
Also, the silicon abundance structure confirms
earlier results published by Hatzes (1995). 

Furthermore, there are strong 
indications that CU Virginis has slowed down its rotation
rate abruptly in the year 1985 (Pyper et al, 1998). 
This effect has not been found in any other Ap star
and no theoretical explanation can be given at the moment.

Nevertheless, the Doppler imaging results for CU Virgins 
are a good demonstration of how important it is
to obtain the surface distribution of different elements
for the same star. CU Virginis is one of the very few objects for 
which a correlation of the abundance structures
and the magnetic field geometry can be given with high significance. 
Furthermore, it is planned to obtain the abundance distributions
for elements like C, N, O and some rare earth species. This
should complete the picture and may lead to a detailed
theoretical modelling of diffusion processes interacting with the
magnetic field of this star.

\end{document}